\documentclass[preprint,showpacs,preprintnumbers,amsmath,amssymb,superscriptaddress]{revtex4}

\usepackage{amssymb}

\usepackage{graphicx}
\usepackage{epsfig}
\usepackage{color}

\usepackage{amsmath,amssymb}

\newcommand{\be}{\begin{equation}}
\newcommand{\ee}{\end{equation}}
\newcommand{\ba}{\begin{eqnarray}}
\newcommand{\ea}{\end{eqnarray}}
\newcommand{\ban}{\begin{eqnarray*}}
\newcommand{\ean}{\end{eqnarray*}}
\newcommand{\ket}[1]{\mbox{$ | #1 \rangle $}}
\newcommand{\bra}[1]{\mbox{$ \langle #1 | $}}

\newcommand{\one}{\leavevmode\hbox{\small1\normalsize\kern-.33em1}}

\begin{document}

\title{Coherent control of long-distance steady state entanglement\\in lossy resonator arrays }
%%%%alphabetical listing
\author{Dimitris G. Angelakis}\email{dimitris.angelakis@googlemail.com}
\affiliation{Centre for Quantum Technologies, National University of
Singapore, 3 Science Drive 2, Singapore 117543} \affiliation{Science
Department, Technical University of Crete - 73100, Chania, Crete,
Greece}
\author{Li Dai}
\affiliation{Centre for Quantum Technologies, National University of
Singapore, 3 Science Drive 2, Singapore 117543}
\address{Department of Physics, National University of
Singapore, 2 Science Drive 3 Singapore 117542}
\author{Leong Chuan Kwek}
\affiliation{Centre for Quantum Technologies, National University of
Singapore, 3 Science Drive 2, Singapore 117543}
\affiliation{National Institute of
      Education and Institute of Advanced Studies,
      Nanyang Technological University, 1 Nanyang Walk, Singapore
      637616}

\begin{abstract}
We show that coherent control of the steady-state long-distance
entanglement between pairs of cavity-atom systems in an array of
lossy and driven coupled resonators is possible. The cavities are
doped with atoms and are connected through wave guides, other
cavities or fibers depending on the implementation. We find that the
steady-state entanglement can be coherently controlled through the
tuning of the phase difference between the driving fields. It can
also be surprisingly high in spite of the pumps being classical
fields. For some implementations where the connecting element can be
a fiber, long-distance steady state quantum correlations can be
established. Furthermore, the maximal of entanglement for any pair
is achieved when their corresponding direct coupling is much smaller
than their individual couplings to the third party. This effect is
reminiscent of the establishment of coherence between otherwise
uncoupled atomic levels using classical coherent fields. We suggest
a method to measure this entanglement by analyzing the correlations
of the emitted photons from the array and also analyze the above
results for a range of values of the system parameters, different
network geometries and possible implementation technologies.
%
% The cavity in which the polariton's state is traced out
%serves as a mediator, which is strangely uncorrelated to the other
%two cavities at steady state.
\end{abstract} 

\date{\today}
\maketitle

 Coupled-cavity arrays have recently been proposed
 as a new system for realizing schemes for quantum computation
\cite{angelakis-ekert04} and for simulations of quantum many-body
systems \cite{simulation of many body system}. More recently driven
arrays were considered towards the production of steady-state
polaritonic \cite{two-state} and membrane entanglement \cite{ple-hue-har}  under realistic dissipation
parameters. Also, an analogy with Josephson oscillations was shown
and the many body properties of the driven array have been recently
studied \cite{coherent control of photon emission}.

\begin{figure}[h]
\epsfxsize=.30\textwidth \epsfysize=.20\textwidth
\centerline{\epsffile{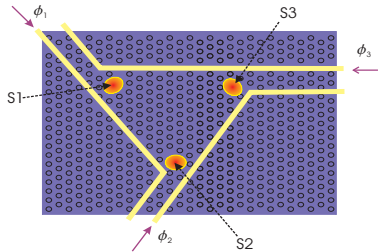}}
%\vspace*{5mm}
\caption{(color online).  Schematic representation of three
interacting cavity-atom systems ($S_{1}$, $S_{2}$, $S_{3}$) setup
based on a possible implementation using photonic crystals (for
illustration purposes only): The connecting wave guides carrying the
driving classical fields with phases $\phi_{1}$, $\phi_{2}$,
$\phi_{3}$ are replaced by fibers or stripline microresonators
for different implementations \cite{pbgs,rest}. The three wave
guides and three driving fields are labeled with the same indices to
the phases $\phi_{1}$, $\phi_{2}$, $\phi_{3}$. } \label{Fig.3-cav}
\end{figure}

In this work we examine for the first time the possibility of
achieving coherent control of the steady-state entanglement
between mixed light-matter excitations (polaritons) generated in
macroscopically separated atom-cavity systems.
  We show explicitly that for a three pumped cavity setup,
which could be realizable in a variety of cavity QED technologies
including photonic crystals, circuit QED, toroidal cavities
connected through fibers and coupled defect cavities interacting
with quantum dots \cite{pbgs,rest}, such control is possible (see
Fig. \ref{Fig.3-cav}). Light from the connecting waveguides/fibers
can directly couple to the photonic modes of the atom-cavity systems
through tunneling or evanescent coupling. In each atom-cavity site
we assume the interaction and the corresponding nonlinearity to be
strong enough to so that either zero or maximally one polariton can
be excited \cite{polariton}.

The Hamiltonian describing the system written in the rotating frame
of the driving lasers is \ba\label{H-original}
H&=&\sum_{i=1}^{3}((\omega_{c,i}-\omega_{d})a_{i}^{\dagger}a_{i}+(\omega_{p,i}-\omega_{d})P_{i}^{\dagger}P_{i})\nonumber\\
&+&\sum_{i=1}^{3}J_{i}(a_{i}^{\dagger}(P_{i}+P_{i+1})+a_{i}(P_{i}^{\dagger}+P_{i+1}^{\dagger}))\nonumber\\
&+&\sum_{i=1}^{3}(\alpha_{i}e^{i\phi_{i}}a_{i}^{\dagger}+\alpha_{i}e^{-i\phi_{i}}a_{i}),\ea
where the first line is the free Hamiltonian of the wave guides and
cavities, with $a_{i}^{\dagger}$, $a_{i}$ the field operators of the
single-mode wave guides.  $P_{i}^{\dagger}$ ($P_{i}$) the operators
describing the creation (annihilation) of a mixed atom-photon
excitation (polariton) at the $i$th cavity-atom system
($P_{4}\triangleq P_{1}$) \cite{polariton} . The second line
describes couplings between cavities and wave guides, with
$\omega_{c,i}$, $\omega_{p,i}$ and $\omega_{d}$ the frequencies of
$i$th waveguide mode, the polariton in $i$th cavity and the driving
fields respectively, and $J_{i}$ is the coupling strength between
the photon mode in the $i$th waveguide and the adjacent two
polaritons. The third line describes the classical driving of the
wave guides, where $\alpha_{i}$ is proportional to the amplitude of
the $i$th driving field with $\phi_{i}$ being its phase.

The polaritons and waveguide modes are assumed to decay with rates
$\gamma$ and $\kappa$ respectively.
% We also assume an infinite-bandwidth limit where $\kappa\gg\delta_{i}$ and
%that the Stark frequency shift can be made to cancel by an
%appropriate choice of $\delta_{i}$.
The master equation for the polaritonic density matrix, after
tracing out the degree of freedom of the waveguide photons
\cite{polariton,method for trace}, is \ba\label{eq-eff}
\dot{\rho}=-i[H_{\mbox{\rm
eff}},\rho]&+&\sum_{i=1}^{3}(\Gamma_{i-1}z_{i-1}+\Gamma_{i}z_{i})F_{i,i}^{P}\rho\nonumber\\&+&\sum_{i=1}^{3}\Gamma_{i}(F_{i,i+1}^{P}\rho+F_{i+1,i}^{P}\rho)\,,\ea
with $\displaystyle H_{\mbox{\rm
eff}}=\sum_{i=1}^{3}(\Gamma_{i}y_{i}P_{i}^{\dagger}P_{i+1}$
+$\Gamma_{i}x_{i}(P_{i}^{\dagger}+P_{i+1}^{\dagger}))+h.c.\,$,$\vspace*{1mm}$
where $h.c.$ denotes the Hermitian conjugation of its previous
summation. $F_{i,j}^{P}(\rho)=2P_{i}\rho
P_{j}^{\dagger}-P_{i}^{\dagger}P_{j}\rho-\rho
P_{i}^{\dagger}P_{j}\,$, $\displaystyle
\Gamma_{i}=J_{i}^{2}\kappa/(\kappa^{2}+\Delta_{i}^{2})$,$\vspace*{2mm}$
$x_{i}=\alpha_{i}e^{i\phi_{i}}(\Delta_{i}-i\kappa)/(J_{i}\kappa)$,
$y_{i}=\Delta_{i}/\kappa$,
$\Delta_{i}=\omega_{c,i}-(\omega_{p,i}+\omega_{p,i+1})/2$,
$\omega_{p,4}\triangleq \omega_{p,0}$,
$z_{i}=1+\gamma/(2\Gamma_{i})$, $\Gamma_{0}\triangleq\Gamma_{3}$ and
$z_{0}\triangleq z_{3}$. It can be seen from Eq. (\ref{eq-eff}) that
the couplings and detunings between the wave guide and its adjacent
two polaritons induce an effective interaction between them given by
$\Gamma_{i}y_{i}$ (see $H_{\textrm{eff}}$). The driving on the wave
guides is equivalently transferred to the driving on the polaritons
($\Gamma_{i}x_{i}$ in $H_{\textrm{eff}}$), which decay with rates
$\Gamma_{i-1}z_{i-1}+\Gamma_{i}z_{i}=\Gamma_{i-1}+\Gamma_{i}+\gamma$.
Since $\Gamma_{i}$ is related to $\kappa$, the polaritons
effectively have two channels of decay. They decay directly to the
outside with $\gamma$ and also through the coupling $J_{i-1}$ or
$J_{i}$ ($J_{0}\triangleq J_{3}$) to the adjacent two leaky wave
guides (who also decay by $\kappa$). Note that the second channel
also mixes the polaritons' operators, as can be seen in the second
line of Eq. (\ref{eq-eff}). This mixing is actually an essential
factor for the entanglement creation among polaritons (the other two
essential factors are the interactions among polaritons and the
driving on them).

\begin{figure}
\epsfxsize=.50\textwidth \epsfysize=.50\textwidth
\centerline{\epsffile{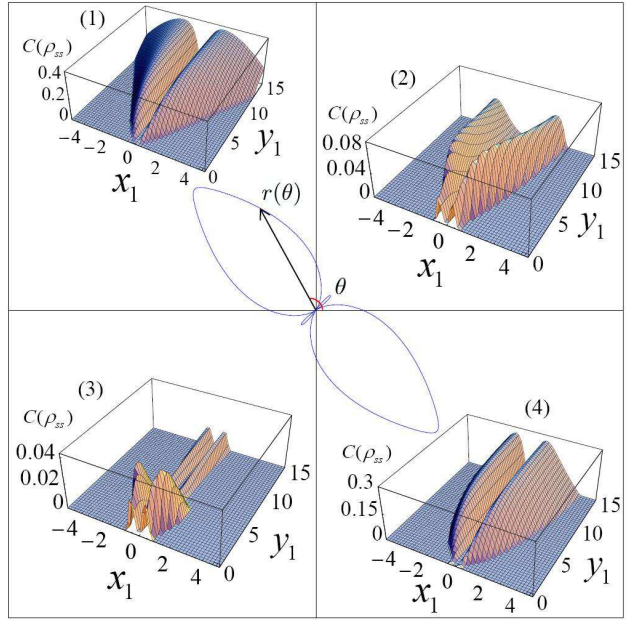}} \vspace*{0mm} \caption{(color
online). Polar plot $r$($\theta$) of the maximum possible
concurrence as the ratio of $x_{1}$ and $x_{3}$ is varied (fix
$y_{1}=y_{3}=15$). $r=C(\rho_{ss})$, tg$\displaystyle
\theta=\frac{x_{3}}{x_{1}}$ and sign($x_{1}$)=sign(cos$\theta$). The
insets (1)-(4) are the 3D plots of $C(\rho_{ss})$ as a function of
$x_{1}$ and $y_{1}$ ($=y_{3}$) with $\displaystyle
\frac{x_{3}}{x_{1}}$ fixed to be $-1$, 1, 0.5, $-0.5$, respectively.
} \label{Fig.x1x3c}
\end{figure}

We can now derive the steady state $\rho_{ss}$ by requiring that
$\dot{\rho_{ss}}=0$ in Eq. (\ref{eq-eff}). This is done numerically
due to the large number of coupled equations
involved\cite{polariton}. Next, for a total three-polariton density
matrix, we trace out the polaritonic degree of freedom of cavity 1
and calculate the polaritonic entanglement between cavity 2 and 3
using the concurrence as a measure. The concurrence of a two-qubit
density matrix $\rho$ is defined \cite{Woot} as max$\{0,
\lambda_{1}-\lambda_{2}-\lambda_{3}-\lambda_{4}\}$, where
$\lambda_{i}$'s are, in decreasing order, the nonnegative square
roots of the moduli of the eigenvalues of $\rho.\tilde{\rho}$ with
$\tilde{\rho}=(\sigma_{1}^{y}\otimes\sigma_{2}^{y}).\rho^{*}.(\sigma_{1}^{y}\otimes\sigma_{2}^{y})$
and $\rho^{*}$ is the complex conjugate of $\rho$. The concurrence
$C(\rho_{ss})$ is effectively a function of the parameters
$x_{i}$\,, $y_{i}$ and $z_{i}$. We perform a numerical optimization
of $C(\rho_{ss})$ by varying these parameters and find that
$C(\rho_{ss})$ is larger when $\Gamma_{2}\ll\Gamma_{1}=\Gamma_{3}$.
For instance, if we assume $\Delta_{1}$ = $\Delta_{2}$ =
$\Delta_{3}$ = 1.5$\times 10^{14}Hz$ and $\kappa$ = $10^{13}Hz$, the
maximum concurrence can reach $0.402$ at $x_{1}$ = $-x_{3}$ = 1.82,
$x_{2}$ = 0, $y_{1}$ = $y_{2}$ = $y_{3}$ = $15$, $z_{1}$ = $z_{3}$ =
$1.113$, and $z_{2}$ = $114$. These correspond to field amplitudes
$\tilde{\alpha}_{1}$ = $\tilde{\alpha}_{3}$ =
$1.215$$\times$$10^{3}$, and couplings $G_{1}$ = $-G_{3}$ =
1.0$\times$$10^{8}Hz$, $G_{2} = \tilde{\alpha}_{2}$ = 0,
($\alpha_{i}=G_{i}\tilde{\alpha}_{i}$), $\phi_{1}=0$,
$\phi_{3}=\pi$, $\gamma$ = $10^{8}Hz$, $J_{1}$ = $J_{3}$ =
1.0$\times$$10^{12}Hz$, $J_{2}$ = 3.16$\times$$10^{10}Hz$. The
effective dissipation rates appearing in the initial Master equation
(Eq. (\ref{eq-eff})) are $\Gamma_{1}$ = $\Gamma_{3}$ =
4.42$\times$$10^{8}Hz$ and $\Gamma_{2}$ = 4.41$\times$$10^{5}Hz$.
These values are consistent with the parameters used in current or
near-future technologies\cite{pbgs,rest}. Fig. \ref{Fig.x1x3c} shows
a plot of the maximum possible concurrence for the polaritonic
entanglement of cavity 2 and cavity 3 when the ratio between $x_{1}$
and $x_{3}$ is varied, with $\Gamma_{1}=\Gamma_{3}$,
$\Gamma_{2}=10^{-3}\Gamma_{1}$, $y_{1}=y_{3}=15$, $z_{1}=z_{3}=1.01$
and $z_{2}=11$. Note that since
$\Gamma_{2}\ll\Gamma_{1}=\Gamma_{3}$, the variation of $x_{2}$ and
$y_{2}$ does not significantly change the value of the concurrence.
It can be seen in Fig. \ref{Fig.x1x3c} that $C(\rho_{ss})$ in the
case when $x_{1}$ and $x_{3}$ have opposite signs is larger than
when they have same signs. $C(\rho_{ss})$ reaches a maximum of 0.417
when $x_{3}=-x_{1}$, i.e. the first and third driving fields have
equal intensity but opposite phases. We also note here that the
relation $\Gamma_{2}\ll \Gamma_{1}=\Gamma_{3}$ indicates that the
coupling between the two cavities in question is much weaker than
the coupling between each one and the third cavity. Also the state
of the polariton in cavity 1 for the maximum entanglement point is
%$\rho_{1}=\mbox{\rm
%Tr}_{2,3}[\rho]=0.99\ket{0}\bra{0}+0.01\ket{1-}\bra{1-}\approx
%\ket{0}\bra{0}$, i.e. that the state for the polariton in cavity 1
found to be almost a pure state at ground energy level and therefore
almost uncorrelated to the polaritons in cavity 2 and 3. Thus, the
total density matrix $\rho\approx
\ket{\textrm{ground}}\bra{\textrm{ground}}\otimes \rho_{2,3}$.
Although this result initially looks counter-intuitive, it can be
explained as follows: the maximum entanglement between the two
parties, i.e. cavities 2 and 3, in a three-party system, is attained
when the state of the third party, i.e. cavity 1, nearly factorizes
in the combined three-party state. The fact that this is happening
for strong relative couplings of $J_{12}\equiv J_1$ and $J_{13}
\equiv J_3$ compared to $J_{23}\equiv J_2$ is reminiscent of the
behavior of a coherent process taking place. One could dare to
observe an analogy here with  the case of coherently superposing two
initially uncoupled ground states in a $\Lambda$ type quantum system
through an excited state using two classical fields to mediate the
interaction \cite{Scully, EIT-Harris}. % CITE{scully book, EIT-Harris}.

\begin{figure}
\epsfxsize=.5\textwidth \epsfysize=.40\textwidth
\centerline{\epsffile{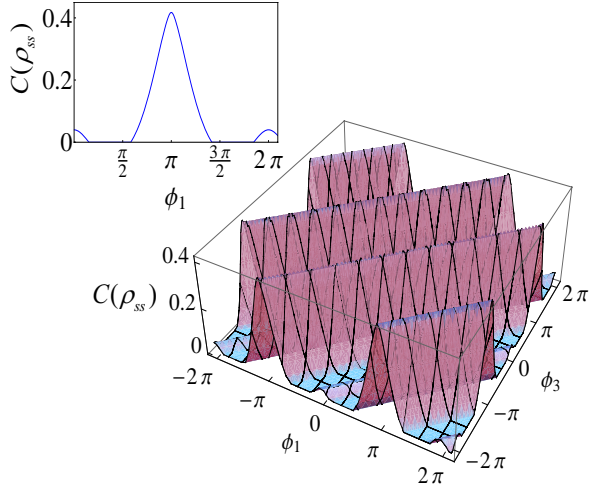}}
%\vspace*{5mm}
\caption{(color online). The concurrence between the polaritons in
cavity 2 and cavity 3 as a function of $\phi_{1}$ and $\phi_{3}$.
$x_{1}=1.67e^{i\phi_{1}}$, $x_{3}=1.67e^{i\phi_{3}}$. When
$\phi_{1}-\phi_{3}=(2k+1)\pi$ (k is an integer), the concurrence
reaches a maximum of 0.417. The upper left figure is the sectional
view at $\phi_{3}=0$.} \label{Fig.phase13}
\end{figure}

The last observation is further justified by observing that
$C(\rho_{ss})$ is larger when the first and third driving fields
have opposite phases. In Fig. \ref{Fig.phase13} we plot
$C(\rho_{ss})$ against the phases of driving fields with
$z_{1}=z_{3}=1.01$ and $z_{2}=11$. When the phase difference is
$\phi_{1}-\phi_{3}=(2k+1)\pi$ (k is an integer), we get again a
maximum of 0.417. For general phase relations, an oscillatory
behavior characteristic of the expected coherent effect takes place.
In simple words, when the two fields are completely out of phase the
entanglement is maximized whereas at phase difference $\pi/2$, the
two polaritons are completely disentangled. In all other cases, the
amount of entanglement lies somewhere in between.

\begin{figure}
\epsfxsize=.25\textwidth \epsfysize=0.20\textwidth
\centerline{\epsffile{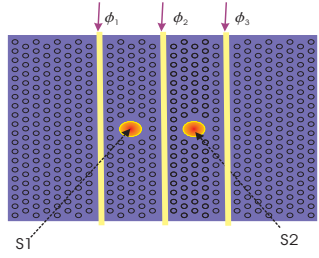}} \caption{(color online).
Schematic diagram of the two cavity-atom systems in which there are
three wave guides carrying the three respective classical laser
fields. Note that each waveguide carrying classical fields can also
be replaced by fibers or stripline microresonators for
implementation technologies \cite{rest}.} \label{Fig.2-cav}
\end{figure}

\begin{figure}
\epsfxsize=.5\textwidth \epsfysize=0.4\textwidth
\centerline{\epsffile{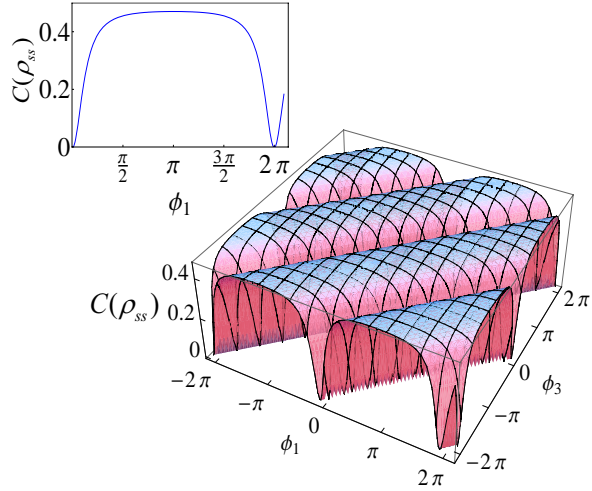}}
%\vspace*{5mm}
\caption{(color online). The concurrence between two cavities
-Fig.\ref{Fig.2-cav}- as a function of $\phi_{1}$ and $\phi_{3}$.
$x_{2}=y_{2}=0$, $x_{1}=5e^{i\phi_{1}}$,$x_{3}=5e^{i\phi_{3}}$,
$\Gamma_{1}=\Gamma_{3}=1.316\times 10^{8}$ and $\Gamma_{2}=10^{10}$.
When $\phi_{1}-\phi_{3}=(2k+1)\pi$ (k is an integer), the
concurrence reaches a maximum of 0.470. The upper left figure is the
sectional view at $\phi_{3}=0$.} \label{Fig.phase23}
\end{figure}

In Fig. \ref{Fig.2-cav}, we study the case of three wave guides
coupled to two cavity-atom systems. Here we analyze the polaritonic
entanglement between cavity 2 and 3 (relabeled as $S_{1}$ and
$S_{2}$ in Fig. \ref{Fig.2-cav}). The optimization of this
entanglement gives similar values of the parameters like the ones
used
 above except that the values for $\Gamma_{i}$ are reversed, i.e.
 $\Gamma_{2}\gg\Gamma_{1}=\Gamma_{3}$; however,
 the concurrence can reach a maximum of 0.47. Again the
dependence $\phi_{1}-\phi_{3}=(2k+1)\pi$ (k is an integer) is
apparent (see Fig. \ref{Fig.phase23}). However, if we compare the
 insets in Fig. \ref{Fig.phase13} and Fig. \ref{Fig.phase23} for the
 cross-sectional plots of the concurrence for $\phi_{3}=0$, we
 see that the plot in Fig. \ref{Fig.phase13} has a narrower peak
 whereas the plot in Fig. \ref{Fig.phase23} is broader. This implies
 that the maximum concurrence for configuration in Fig.
 \ref{Fig.2-cav} is substantially more stable against variation in
 the phases $\phi_{1}$ and $\phi_{3}$ than that in Fig.
 \ref{Fig.3-cav}.

\begin{figure}[t]
\epsfxsize=.5\textwidth \epsfysize=0.35\textwidth
\centerline{\epsffile{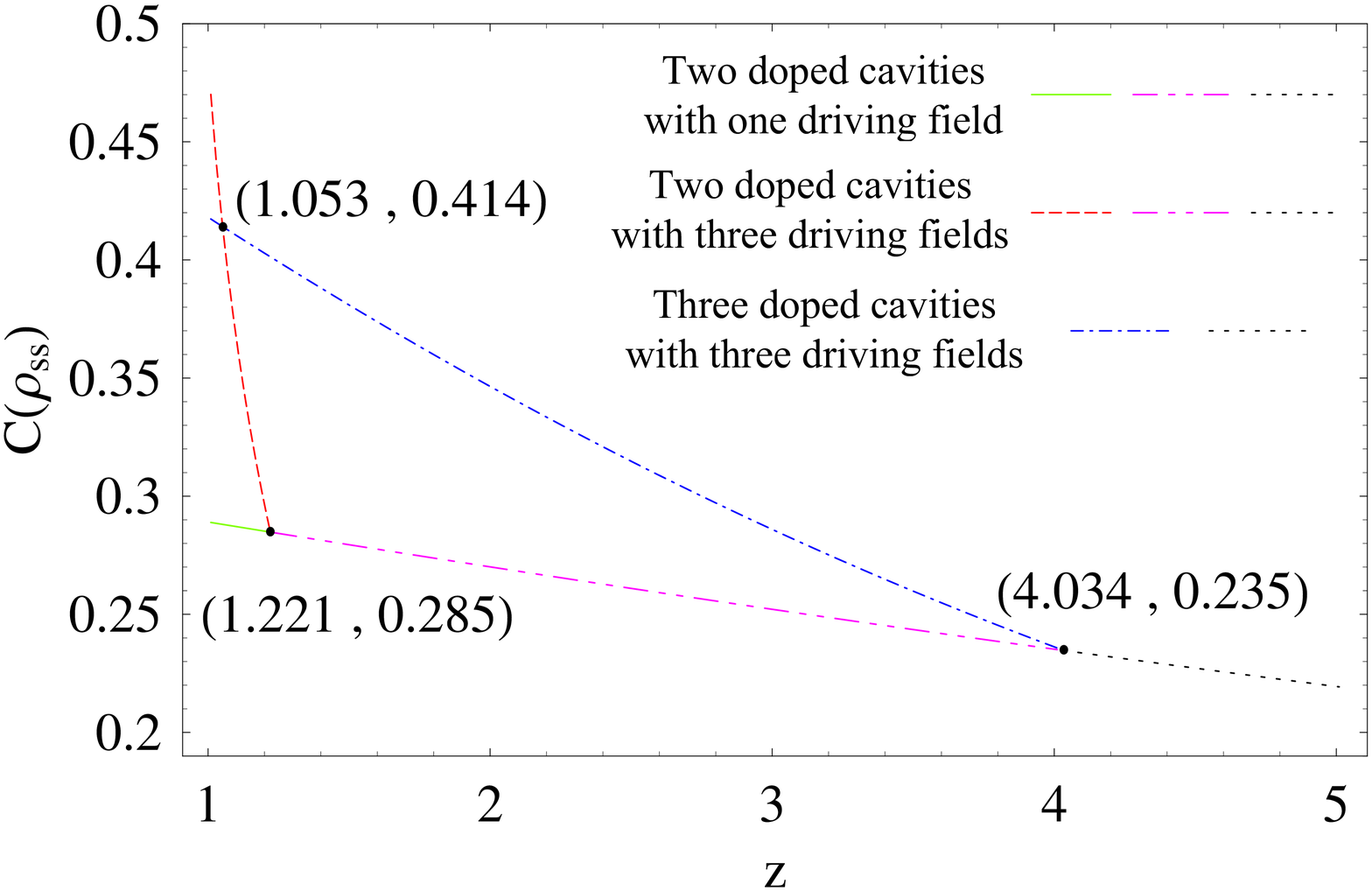}}
%\vspace*{5mm}
\caption{(color online). The maximum concurrence versus $z$ in three
configurations: (i), two cavities with a single driven waveguide in
Ref. \cite{two-state}; (ii), two cavities with three driven wave
guides as in Fig. \ref{Fig.2-cav}; (iii), three cavities with three
driven wave guides as in Fig. \ref{Fig.3-cav} ($z_{1}=z_{3}=z$,
$z_{2}=10^{3}(z_{1}-1)+1$). The solid/dashed line is for
configuration (i)/(ii) when $1<z<1.221$. The dash dot line is for
configuration (iii) when $1<z<4.034$. The double dot dash line is
for configuration (i) and (ii) when $1.221<z<4.034$. The dot line is
for all the three configurations when $z>4.034$.}
\label{Fig.3-cross}
\end{figure}

At this juncture, it is worth emphasizing that we now have three
different configurations for comparisons: (i) two cavities with a
single driven wave guide in Ref. \cite{two-state}; (ii) two cavities
with three driven wave guides as in Fig. \ref{Fig.2-cav}; (iii)
three cavities with three driven wave guides as in Fig.
\ref{Fig.3-cav}. Numerical optimization involving more than three
doped-defect cavities do not seem to increase the polaritonic
entanglement between any two cavities. Therefore, the above three
configurations should be optimal for two-qubit entanglement,
corresponding to different values of the dissipation rates
parametrized in $z$. As shown in Fig. \ref{Fig.3-cross}, when $z$
ranges from 1 to 1.221, the maximum concurrence for configuration
(ii) decreases rapidly from 0.48 to 0.285. This rapid decrease
indicates that although configuration (ii) can reach higher
entanglement than configuration (i), yet it is more fragile to the
dissipation of the environment parametrized by $\gamma$ in $z$). In
comparison, the three-cavity setup is more robust against the
increase of dissipation (only when $z\gtrsim 4.03$, its maximum
concurrence drops to be the same to that for configuration (i)).
Therefore, we conclude that cavity 1 in Fig. \ref{Fig.3-cav} not
only coherently mediates between cavities 2 and 3, but it also
stabilizes the amount of entanglement between the two cavities.

\begin{figure}[h]
\epsfxsize=.4\textwidth \epsfysize=0.25\textwidth
\centerline{\epsffile{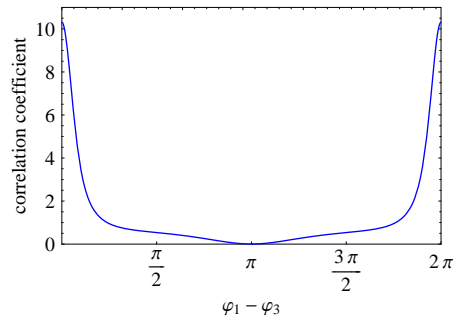}}
%\vspace*{5mm}
\caption{(color online). (The cross-correlation coefficient
$\displaystyle \frac{\langle P^\dagger_2 P_{2} P^\dagger_3
P_{3}\rangle}{\langle P^\dagger_2 P_{2} \rangle \langle P^\dagger_3
P_{3} \rangle}$ for the three-cavity scheme: the minimum value in
the cross-correlation coefficient corresponds to maximum concurrence
between the cavities.} \label{Fig.acphase123}
\end{figure}

One could try to employ entanglement witnesses to detect this
entanglement \cite{witness}. A witness could be constructed from the
density matrix corresponding to the maximum value of the concurrence
\cite{two-state} and one could measure the witness along the
corresponding spin directions. In coupled-cavity systems to
implement the necessary effective spin measurements we can use the
usual atomic state measurement techniques employing external laser
tuned to the corresponding polaritonic levels \cite{coherent control
of photon emission,polariton}. In these measurements the
correlations between the polaritons are transferred to emitted
photons and can thus be detected by analyzing the fluorescent photon
spectrum. In the following we plot the cross-correlation coefficient
$\displaystyle \frac{\langle P^\dagger_2 P_{2} P^\dagger_3
P_{3}\rangle}{\langle P^\dagger_2 P_{2} \rangle \langle P^\dagger_3
P_{3} \rangle}$ for the three-cavity scheme in Fig. \ref{Fig.3-cav}
as a function of the phase difference between the driving field 1
and 3 (Fig. \ref{Fig.acphase123}). The plot is consistent with the
concurrence plot in Fig. \ref{Fig.phase13}. What we observe is that
 when the polaritons are highly entangled the emitted photons
 come in bunches from each polariton emitter (we note
 here that the polaritons are continuously pumped).

In this work, we have shown that long-distance steady state entanglement in a
lossy network of driven light-matter systems can be coherently
controlled through the tuning of the phase difference between the
driving fields. This entanglement could be measured by analyzing
the spectrum of the photons  emitted from the cavities.
  We also found that there exist two optimal setups for generating
maximum available entanglement between two coupled cavity systems
depending on the level of dissipation in the system. Finally,
surprisingly enough, in a closed network of three-cavity-atom
systems
 the maximum of entanglement for any pair is
achieved even when their corresponding direct coupling is much
smaller than their couplings to the third party. This effect is
reminiscent of coherent effects found in quantum optics that
coherent population transfers between otherwise uncoupled levels
through a third level using two classical coherent fields.

Acknowledgment -
We would like to acknowledge financial support by the National Research Foundation \&
Ministry of Education, Singapore. We would also like to thank Stefano Mancini for helpful comments.
D.G.A. and L.C.K would like to thank the Centro de Ciencias de Benasque ``Pedro Pascual" for the hospitality at the Benasque Workshop on Quantum Information
where part of this work was done.

%%%%%%%%%%%%%%%%%%%%%%%%


\begin{thebibliography}{99}
\bibitem{polariton}
There are several ways to form polaritons through atom-photon
interactions. For instance, a two-level atom interacting with cavity
photons can exihibit nonlinear energy spectrum and in the
photon-blockade regime, only the lowest two levels needs to be
considered which form the two levels for excitations of poaritons.
The ground state of the polariton is $|g,0\rangle$ and the excited
state is the lowest dressed state
$(|e,0\rangle-|g,1\rangle/)\sqrt{2}$. These two states on resonance
are separated by $\omega_{pol}=\omega_{0}-g$ with $\omega_{0}$ the
frequency of the uncoupled atomic levels/photon and $g$ the
atom-photon coupling strength. There can also be alternatives
involving 4-level atoms in each cavity interacting with the photon
through the usual Jaynes-Cummings interaction. See also Ref.
\cite{Scully,EIT-Harris}. In the above regime if one calculates the
commutator between the polaritons, a mixed relation will be found
which only in the limit of large detunings and/or small couplings
leads to bosonic one. This is the limit where the polaritons are
mostly comprised of photons.
\bibitem{method for trace}
Here, we assume weak pumps: $\alpha_{i} \le J_{i} \ll \kappa$. The
method for tracing out the degree of freedom of waveguide modes is
similar to the one in Ref. \cite{two-state}.
\end{thebibliography}
\end{document}